\documentclass[9pt,twocolumn,twoside]{opticajnl}
\journal{opticajournal} 

\setboolean{shortarticle}{true}

\usepackage{cleveref}
\usepackage{lineno}
\nolinenumbers 
\usepackage{multirow}
\usepackage{xcolor}
\usepackage{subcaption}
\usepackage{ulem}
\usepackage{setspace}

\title{Vignetting Effects: a Tool to Characterize a Fourier Ptychographic Microscope}

\author[1,*]{John Meshreki}
\author[1]{Syed Muhammad Kazim}
\author[1]{Jan Philipp Schneider}
\author[1]{Michael Moeller}
\author[1]{Ivo Ihrke}

\affil[1]{University of Siegen, Germany.}

\affil[*]{john.meshreki@uni-siegen.de}

\newcommand{\magobj}{M_{\textit{Obj}}}
\newcommand{\magled}{M_{\textit{LED}}}
\newcommand{\diamvc}{d_{\textit{VC}}}

\newcommand{\vect}[1]{\boldsymbol{#1}}

\newcommand{\obsvectgeom}{\vect{\phi}_{\textit{g}}}
\newcommand{\sysvectgeom}{\vect{\theta}_{\textit{g}}}
\newcommand{\sysvectgeomw}{\hat{\vect{\theta}}_{\textit{g}}}
\newcommand{\obsvectcoh}{\vect{\phi}_{\textit{c}}}
\newcommand{\sysvectcoh}{\vect{\theta}_{\textit{c}}}

\newcommand{\dist}[2]{\overline{{#1}{#2}}}

\newcommand{\simgeom}{f_{\textit{geom}}}
\newcommand{\simwave}{f_{\textit{wave}}}

\newcommand{\argmin}{\textrm{argmin}}

\begin{abstract}
  Fourier Ptychographic Microscopy (FPM) is a recent technique to
  overcome the diffraction limit of a low numerical aperture (NA) objective lens by
  algorithmic post-processing of several lower resolved images.
  It can increase the space-bandwidth product of an optical system by
  computationally combining images captured under different
  illumination conditions.
  Vignetting determines the spatial extent of the bright field and dark
  field regions in the captured images that contain information about
  low and high frequency image content, respectively.
  State-of-the-art analyses treat vignetting as a nuisance that
  needs to be reduced or excluded from algorithmic consideration by
  means of ad-hoc decision rules~\cite{pan_2019_vignetting}.
  In contrast, this work investigates vignetting effects as a tool to
  infer a range of properties of the optical system.
  To achieve this, we characterize the individual system components
  of the experimental setup and compare experimental data to both,
  geometrical and wave optical simulations.
  We demonstrate that using vignetting as an analytical tool enables
  the modeling of the geometric and coherence properties of the
  optical system as evidenced by the good agreement between our
  simulation and experiment.
  Moreover, our work investigates pupil aberrations in the FPM setup
  and enables their partial characterization, despite not yet
  encompassing all aspects.
\end{abstract}

\setboolean{displaycopyright}{false} 

\begin{document}

\maketitle

\begin{figure}[ht]
  \centering
  \fbox{\includegraphics[width=0.9\columnwidth]{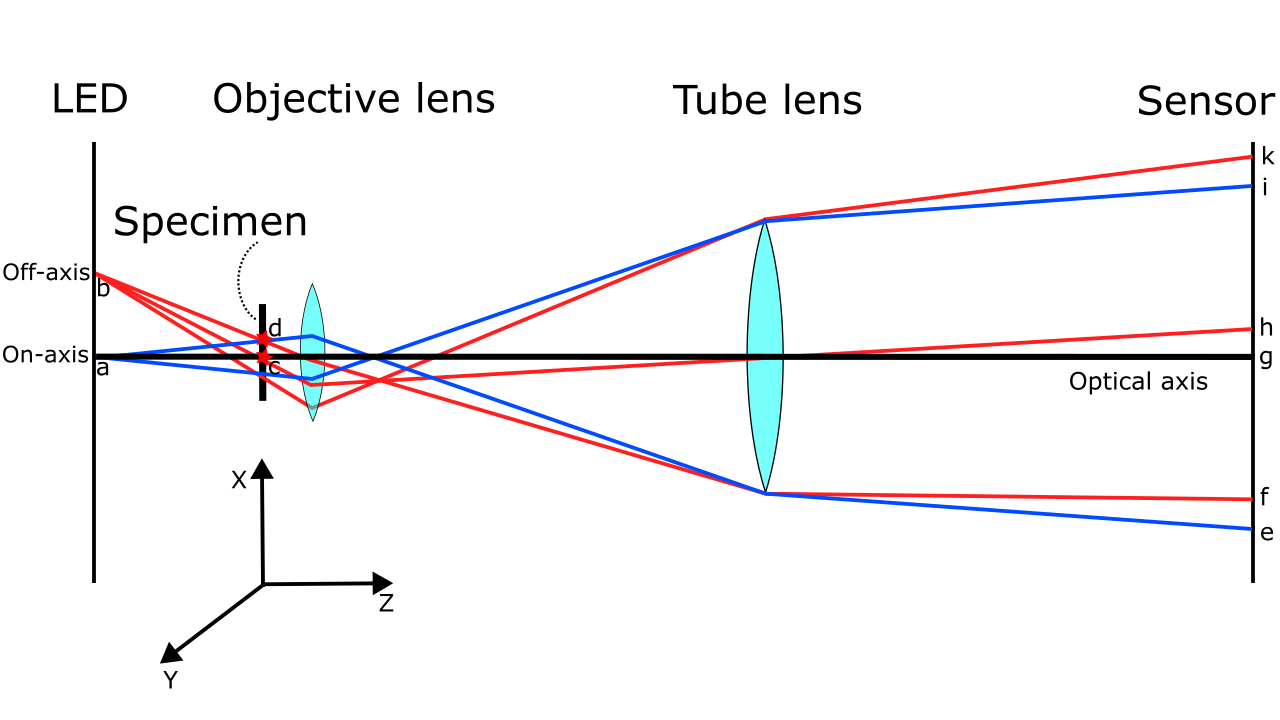}}
  \caption{A ray diagram of the FPM setup with $X$-$Y$ planes
    representing the LED illumination, specimen, objective lens, tube
    lens and sensor.  The coordinate system of the setup is also
    displayed.  The blue and red rays correspond to the on-axis and
    off-axis cases, respectively. Image regions within a cone of light
    are referred to as bright field, whereas regions outside of it are
    dark field regions.}
  \label{fig:sketch}
\end{figure}

  FPM is an innovative method for achieving highly resolved 2D and 3D
  reconstructions~\citep{zheng_2013_widefield, tian2014boe,
    tian2015oa} with a wide spectrum of applications in industry and
  biological and medical fields~\citep{zheng2021nrpa, konda2020oe,
    pan2020rpp}.
  This is due to its efficiency in obtaining a high spatial bandwidth
  beyond the diffraction limit of the objective lens formulated by Ernst
  Abbe~\cite{abbe:1873}.
  The FPM setup comprises an LED array, an objective lens, a tube
  lens, and a camera sensor to record the captured images, as can be
  seen in~\Cref{fig:sketch}.
  By turning on one of the LEDs, it sends an electromagnetic field
  which interacts with the specimen placed in front of it and the exit
  field continues to propagate through the two lenses which magnify
  the image and a final image is recorded at the sensor plane.
  In conventional microscope settings, with a constant space-bandwidth
  product, there is a trade-off between capturing high resolution
  information (high-NA) and a wide Field of View (FoV, low-NA
  setting).
  In contrast, FPM typically uses a low NA objective, capturing
  multiple low resolution, but wide FoV images.
  Algorithmic post-processing then computes high-resolution, wide-FoV
  images, i.e.  the space-bandwidth product is increased beyond that
  of the physical system.
  High-quality reconstructions require the knowledge of 1)
  first-order geometrical properties, 2) coherence properties, and
  3) aberrations of the optical system~\cite{yeh2015oe}.  
  Usually, this knowledge is inferred via additional optimization
  goals in the FPM reconstruction procedure~\cite{tian2014boe}.
  Since the resulting optimization problem is non-linear and
  non-convex, it is, however, uncertain whether accurate values are
  being recovered using this process.

\begin{figure}[ht]
\centering
 \includegraphics[height=0.15\textheight, width=0.5\textwidth]{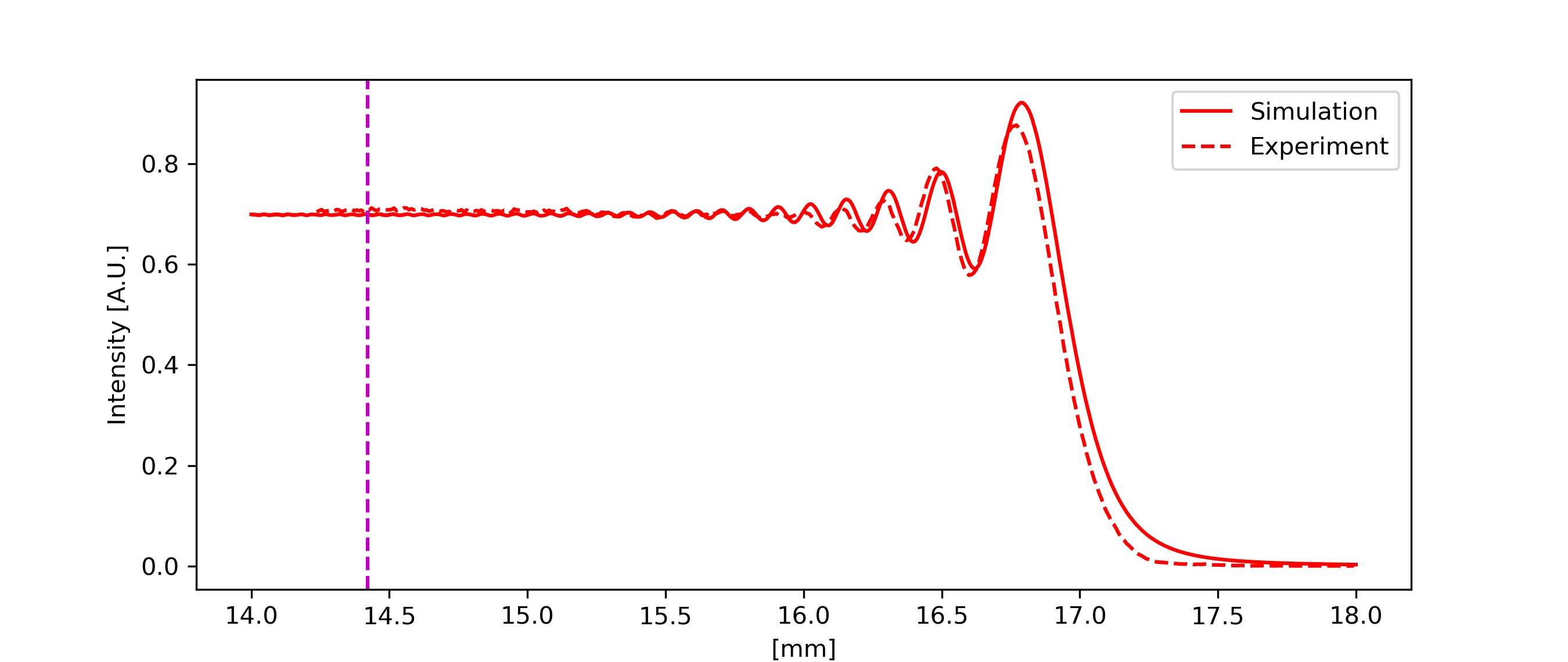}
 \includegraphics[height=0.15\textheight, width=0.5\textwidth]{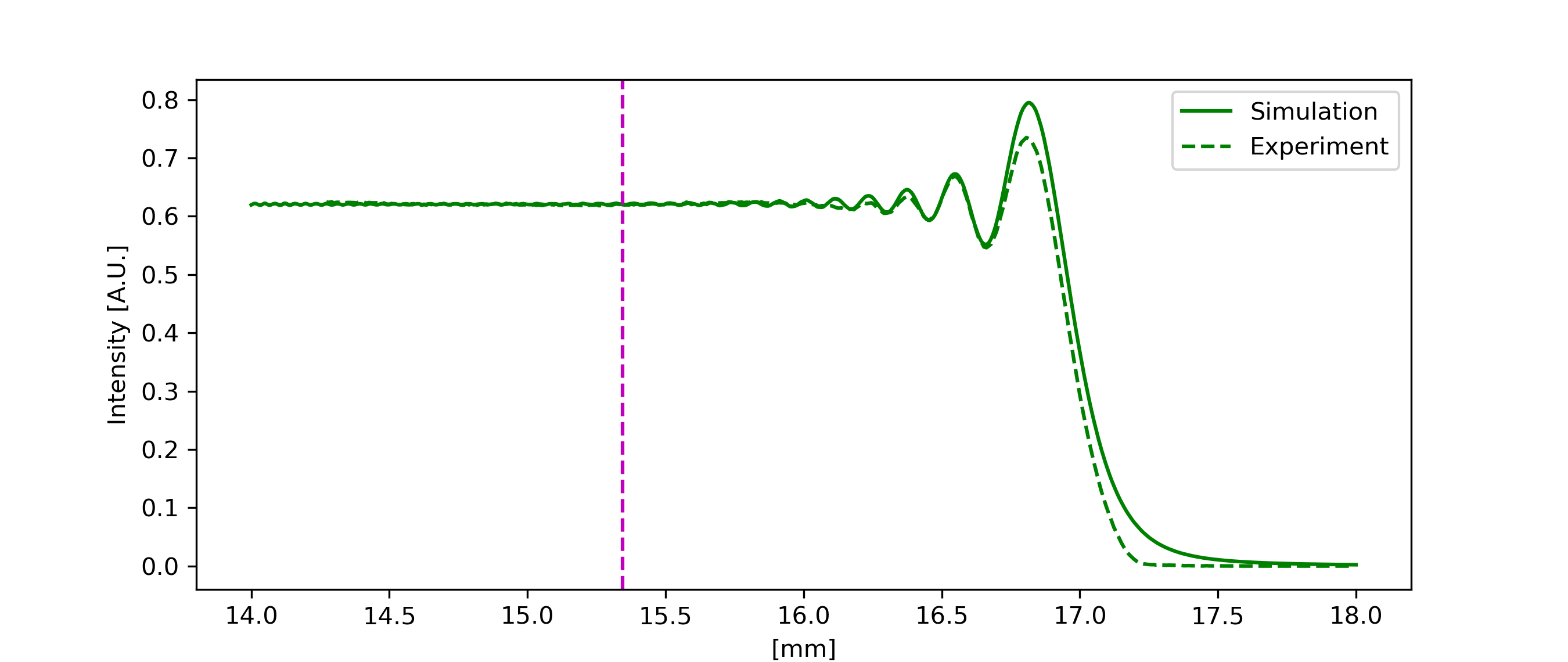}
 \includegraphics[height=0.15\textheight, width=0.5\textwidth]{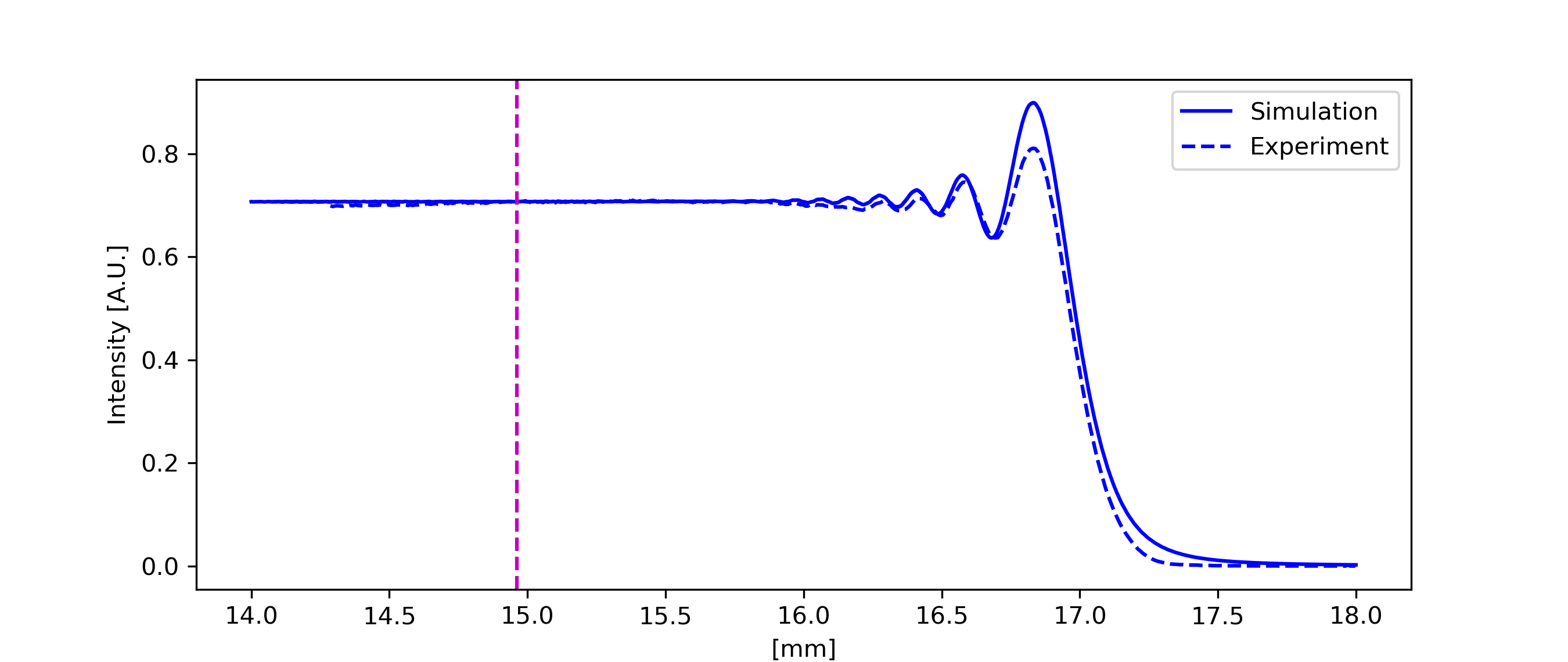}
\caption{VC's profile of the measurements (dashed line) is overlaid to 1D simulation (solid line) for the R, G, and B channels (top to bottom). The vertical lines are the temporal coherence length of the respective wavelength. Notice the peaks extending to the left cease prior to the temporal coherence length due to the influence of the spatial coherence.}
\label{fig:vignet_profile_all_chan}
\end{figure}

  In this work, we therefore aim at identifying system properties
  that can be inferred from direct observation. 
  For this, we adopt a novel perspective on vignetting as a
  characteristic property of the optical system that is easily
  accessible and that carries valuable information about the system
  that can be extracted by careful analysis.
  We show that the geometric properties of a system (1) and its coherence
  properties (2) can be characterized to an accuracy as to enable
  predictive simulation, see for example Fig.~\ref{fig:vignet_profile_all_chan}.
  Moreover, we demonstrate that the presence of 4D aberrations can be
  effectively detected, yet, further studies are required for a
  precise system calibration.
  In addition, vignetting can be used for a suitable alignment of the
  optical system as we show later.

%
  
For the purposes of this letter, we analyze an FPM setup that was
constructed in our lab.  The illumination unit in the experimental
setup employs a programmable Adafruit 32x32 LED array, where each LED
is controlled using an Arduino microcontroller to emit (R,G,B) light
with a spacing of 4mm between each LED.  The setup uses a 10x 0.3NA
Nikon objective lens, combined with a 200mm focal length,
infinity-corrected tube lens (Thorlabs TTL 200).  We use a consumer-grade
camera (Canon 5D mark II) due to its large sensor area as well as a
high resolution of 5,634 × 3,753 pixels with a pixel size of 6.4$\mu$m.
We capture images of the NBS 1963A Pattern using the (R,G,B) modes of
the LEDs.  The camera's large sensor is able to capture almost the
full extent of the unvignetted area, visible as a circle containing the object.

\begin{figure}
    \centering
    \includegraphics[width=0.9\columnwidth]{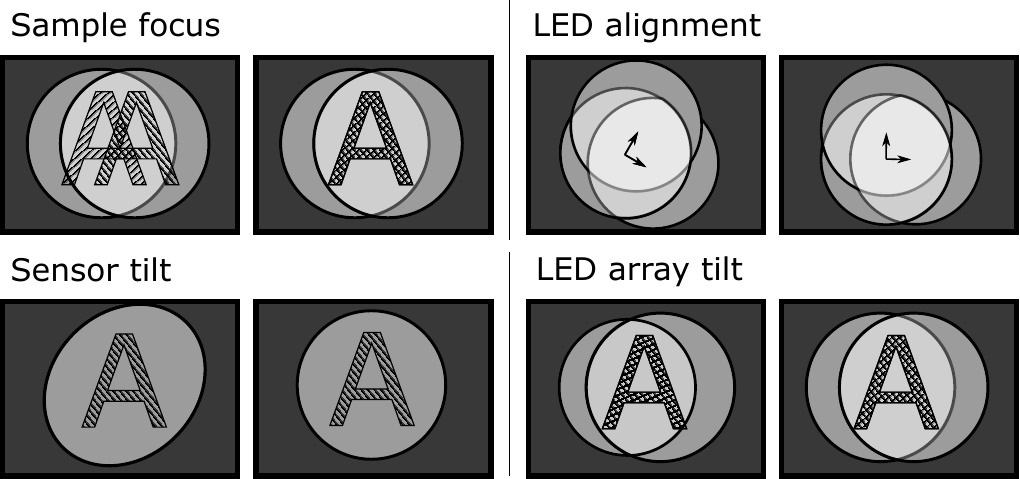}
    \caption{Focusing (top left) and alignment (top right) can be
    finely adjusted activating multiple LEDs. Sensor and LED array tilt
    affect the eccentricity and relative scale of the VCs.}
    \label{fig:alignment}
  \end{figure}

%
  \paragraph{Alignment.}
  A first advantage of studying vignetting lies in its usefulness for a suitable
  alignment of the optical system, see Fig.~\ref{fig:alignment}. If the
  sensor is chosen to be sufficiently large, a bright circle becomes visible
  in the image plane, separating bright field and dark field regions,
  see Fig.~\ref{fig:sketch}.
  In our setup, this vignetting circle (VC)
  is caused by the tube lens restricting the propagation of light for
  on- and near-axis LEDs.
  Assuming a sensor centered on the optical axis, e.g. with the aid of
  an alignment laser, the LED array can be centered by moving its
  center LED such that the vignetting circle is centered on the
  sensor.  This is best done with live view capabilities.
  For further system alignment, it is useful to activate several LEDs,
  e.g. in an ``L'' or cross pattern, see Fig.~\ref{fig:alignment}~(top
  right), since the orientation of the LED array with respect to the
  sensor axis becomes visible and can be adjusted accordingly. In
  addition, an out-of-plane tilt in the LED array manifests as a
  change in the VC's radius (bottom right).
  A second observation is that focusing the microscope with single LED
  illumination is difficult due to the near-coherent illumination and
  the corresponding effectively small NA, causing a very large
  depth of field. Activating multiple LEDs leads to the superposition of several specimen images (illustrated by the ``A'' pattern and assuming it to be
  planar)  when the specimen is out of focus.
  Adjusting the
  focus shifts these copies w.r.t.  each other (top left). Therefore, a good
  focus setting is achieved when the multiple images align on top of
  each other. For 3D samples, a suitable sample plane should be chosen
  as the reference focus in this manner.
  In both the 2D and 3D cases, a larger lateral LED distance from
  the center leads to increased sensitivity.  A sensor tilt is visible
  as an elliptical shape of the VC, Fig.~\ref{fig:alignment}~(bottom
  left).

\paragraph{Geometric Characterization.}
In the following, we aim to match a simulation to observed intensity
images of the vignetting circle.  In a first step, we exploit, as a
second property of the vignetting circle, its sensitivity to the
construction parameters of the optical system. In particular, with
reference to Fig.~\ref{fig:sketch}, the diameter $\diamvc=\dist{e}{i}$
and relative displacement $\magled=\dist{g}{h}/\dist{a}{b}$ of the
vignetting circle add additional information over and above the
standard object magnification $\magobj$ which is measured using a
calibration target.

We formulate the determination of the optimal system parameters as an
optimization problem: Given experimental values of the three
observables, $\diamvc$, $\magled$ and $\magobj$, determine the most
likely construction parameters of the system, i.e. the axial
distances, component diameters, focal lengths, pupil and principal
plane positions as well as lateral LED spacing. Collecting the
geometrical observables in a vector $\obsvectgeom$ and the system
parameters in a vector $\sysvectgeom$, and denoting a first-order ray
optical simulation that is predicting the observables as
$\simgeom(\sysvectgeom)$, we may write
\begin{equation}
  \sysvectgeom^* = \underset{\sysvectgeom}{\argmin}
  \left\lVert \simgeom(\sysvectgeom) - \obsvectgeom \right\rVert_2^2,
  \label{eq:optgeom}
\end{equation}
where $\sysvectgeom^*$ is the vector of geometric system
parameters causing the best fit between simulation and measurement.
We initialize the optimization with the measured parameters ${\sysvectgeom}_0$,
with an estimate of their standard deviations. We then perform a
Monte-Carlo (MC) sampling in the vicinity of the initial guess
${\sysvectgeom}_0$, assuming independent Gaussian errors of the individual
measurements to avoid local minima that may occur in our
over-parameterized system, select the minimum and iteratively
repeat the process with decreased standard deviation values.
We did not observe multiple minima within the measurement
uncertainties which suggests that a gradient descent optimization may be used in the
future.
Using the optimized parameters $\sysvectgeom^*$, we obtain a good agreement between the measurements
and the simulations in all color channels as can be seen in Tab.~\ref{tab:sim_meas_geom_comp}.
Furthermore, such parameters are input to the next step, a fitting of the
coherence properties of the system.

{
\scriptsize
\begin{table}[htbp]
\centering
\begin{tabular}{|c|c|c|c|c|}
\hline
\multirow{2}{*}{$\scriptstyle \textrm{Channel}$} & \multirow{2}{*}{$\scriptstyle \textrm{Type}$} & \multicolumn{1}{c|}{$\scriptstyle d_{VC} \textrm{[mm]}$ } & \multicolumn{1}{c|}{$\scriptstyle M_{obj}$} & \multicolumn{1}{c|}{$\scriptstyle M_{LED}$} \\ 
 &  &  &  &  \\ \hline
\multirow{2}{*}{ $\scriptstyle \textrm{R/G/B}$}    & $\scriptstyle \textrm{Sim.}$       & $\scriptstyle 34.20~\textrm{(all)}$ & $\scriptstyle 9.96~\textrm{(all)} $   & $\scriptstyle 0.32~\textrm{(all)}$ \\ \cline{2-5}
			          & $\scriptstyle \textrm{Meas.}$  & $\scriptstyle 33.89 / 33.97 / 33.99$ & $\scriptstyle 10.16~\textrm{(all)}$ & $\scriptstyle 0.31 / 0.31 / 0.30$ \\ \hline
\end{tabular}
\caption{Comparison of the VC diameter and object and LED magnifications between simulations and measurements for the R, G and B channels.}
\label{tab:sim_meas_geom_comp}
\end{table}
} 

\paragraph{Wave-Optical Characterization.}
Given the geometric properties of the optical system, we upgrade the simulation to the scalar wave regime in order to match the coherence properties of the system between simulation and measurement.
The apertures in the system give rise to
prominent diffraction patterns at the edges of the vignetting circle,
the detailed structure of which is related to a) the geometric system
parameters, most notably the aperture size and the propagation
distance to the sensor, and b) the coherence properties of the light
source.
We exploit this property by analyzing the diffraction pattern for the
center LED in terms of the light source properties, see Fig.~\ref{fig:vignet_profile_all_chan}, and a refinement of
the geometric system parameters, 
to establish the coherence properties of our FPM system.

\begin{figure}[ht]
  \centering
  \fbox{\includegraphics[width=0.45\textwidth]{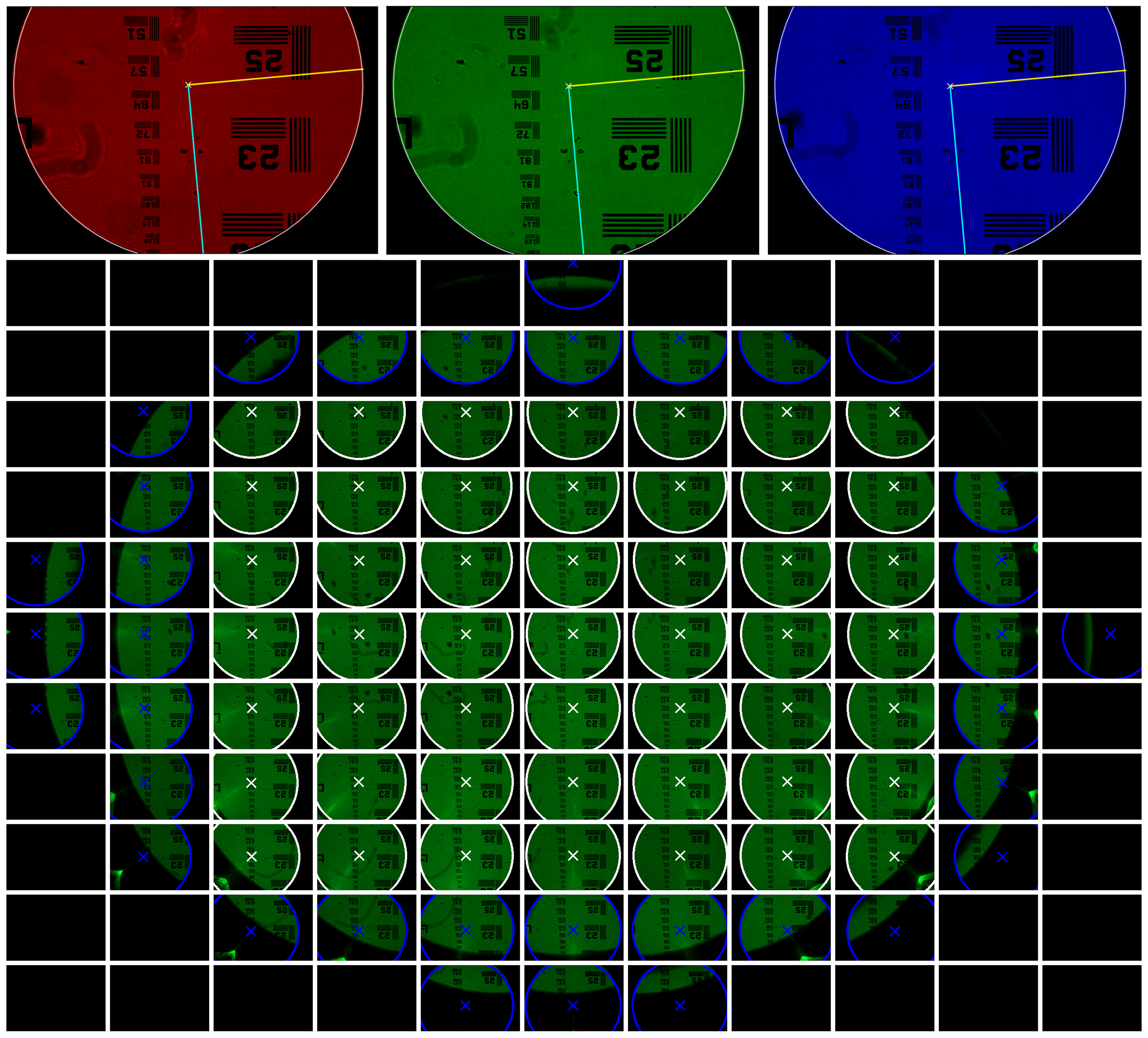}} 
  \caption{Grid of images of the NBS 1963A Pattern illuminated by
    green LEDs and fitted to ellipses. Elliptical contours resulting
    from the VC fits are superimposed on the individual images in
    white/blue. The top row shows the images of the on-axis LED in
    (R,G,B), where the yellow and cyan lines represent the minor and major semi-axes, respectively. }
  \label{fig:grid}
\end{figure}

We measure both the LED spatial emission profile and its emission
spectrum for each color channel , respectively, by conducting a fit
with a Gaussian function to extract the associated means and standard
deviations $\mu_s,\sigma_s$ (spectral) and $\mu_x, \sigma_x$
(spatial), where $\mu_s,\sigma_s$, $\mu_x$ and $\sigma_x$
are determined for each of the (R,G,B) color channels. We capture
images using all three illuminations, see Fig.~\ref{fig:grid}~(top)
and extract an intensity profile along the edge of the VC for each
color channel using image processing: We detect the intensity maxima
along the edge profile, generating a set of points, which we fit with
an ellipse (we observe the geometrical shape to be elliptical
indicating a sensor tilt, see bottom left of
Fig.~\ref{fig:alignment}), establishing its midpoint and its major and
minor axes. By varying the major and minor axes by a common scale
factor, we can generate ellipses that are offset to the inside or outside of the ellipse approximating the intensity maxima. By
averaging across the circumference of each ellipse, we extract a
noise-reduced intensity value for each scale factor, resulting in the
intensity profiles labeled ''Experiment'' in
Fig.~\ref{fig:vignet_profile_all_chan}.

For simulation, we use the angular spectrum method, as described by
Matsushima et al.~\cite{Matsushima:2009}, to accurately model wave
field propagation, which implements Rayleigh-Sommerfeld diffraction
theory without approximation. To simulate the partial coherence of the
LED sources, we perform a MC sampling of the spatial
$(\mu_x=0,\sigma_x)$ and spectral probability distributions
$(\mu_s,\sigma_s)$ determined by our measurements, where
$\mu_x, \sigma_x, \mu_s,$ and $\sigma_s$ are individually treated for the 
(R,G,B) color channels.  We simulate 1000 coherent fields with
uniformly distributed random initial phases that are superposed
incoherently. The individual simulations are initialized with
spherical wavefronts in the object plane and propagated through the
system, treating the microscope objective and the tube lens as thin
lenses with a perfect lens phase profile that maps a divergent
spherical wave from the point source to a spherical wave converging to
its Gaussian image point. For first order properties of the system, we
use the optimized geometric system parameters $\sysvectgeom^*$.
The simulation results in an intensity profile similar to the one
labeled ``Simulation'' in Fig.~\ref{fig:vignet_profile_all_chan}, but
generally not fitting the experimental curve precisely.

We therefore resort to another optimization to fit the intensity
profiles. We vary the light source system parameters
$\sysvectcoh=(\mu^r_s,\sigma^r_s,\sigma^r_x,\mu^g_s,\sigma^g_s,\sigma^g_x,\mu^b_s,\sigma^b_s,\sigma^b_x)$,
where the superscripts indicate the respective values for the color channels.
Moreover, we adjust the $\sysvectgeom^*$ geometric parameters using a new variable
$\sysvectgeomw$, while keeping it close to the previous
result $\sysvectgeom^*$ . Due to the wavy nature of the intensity profiles and the
sensitivity of the curve to the light source parameters it is not
advisable to compare the curves directly. Instead, we opt for
comparing the positions of the first four local intensity maxima (as
seen from the edge) of the profiles,
Fig.~\ref{fig:vignet_profile_all_chan}, collecting them in an
observable vector $\obsvectcoh$. Denoting the wave optical simulation
and subsequent peak fitting as $\simwave( \sysvectcoh;
\sysvectgeom)$, we formulate the optimization as
\begin{equation}
  \sysvectcoh^*, \sysvectgeomw^* =
    \underset{\sysvectcoh, \sysvectgeomw}{\argmin}
    \left\lVert \simwave(\sysvectcoh; \sysvectgeomw) - \obsvectcoh \right\rVert_{2,1} +
    \lambda \left\lVert \sysvectgeomw - \sysvectgeom^* \right\rVert_2^2,
  \label{eq:optcoh}
\end{equation}
where the group L2-norm is taken per local intensity maximum. The
standard deviation in the geometric parameters is iteratively reduced
during optimization. The final fit is shown in
Fig.~\ref{fig:vignet_profile_all_chan}, denoted as simulation.  Note
that the result uses a single set of geometric system parameters
$\sysvectgeomw^*$, whereas the light sources use the respective
$\sysvectcoh^*$. We have therefore obtained a single system
description that can predictively simulate the diffraction behavior of
our system at different wavelengths, as can be seen in
Fig.~\ref{fig:vignet_profile_all_chan}.

\paragraph{Aberrations.}
In a perfectly aligned system, the center LED creates a virtual point
source on the optical axis, illuminating a centered exit pupil
orthogonal to the optical axis, creating a circular image on
a sensor, which is also positioned orthogonally to the axis.
Provided that LEDs form a regular grid, also placed orthogonal to the
axis, illuminating the specimen individually, a similar grid of circles,
of equal sizes, should be observed on the image plane.

\begin{figure}[ht]
  \captionsetup[subfigure]{justification=centering}
  \centering
  \begin{subfigure}{0.23\textwidth}
      \includegraphics[width=\textwidth]{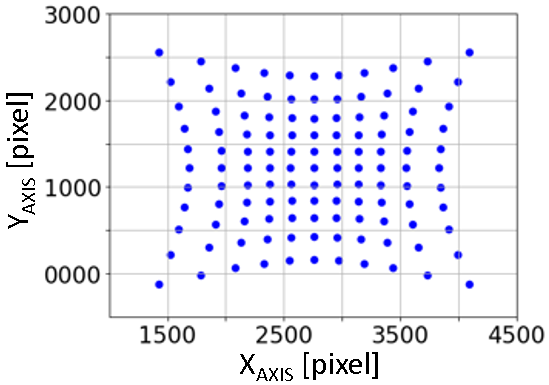}
      \caption{Center Coordinates}
  \end{subfigure}
  \hfill
  \begin{subfigure}{0.21\textwidth}
      \includegraphics[width=\textwidth]{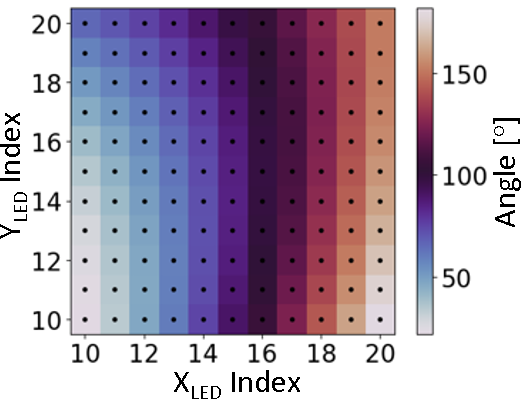}
      \caption{Angles of Rotations}
  \end{subfigure}
  \hfill
  \begin{subfigure}{0.22\textwidth}
      \includegraphics[width=\textwidth]{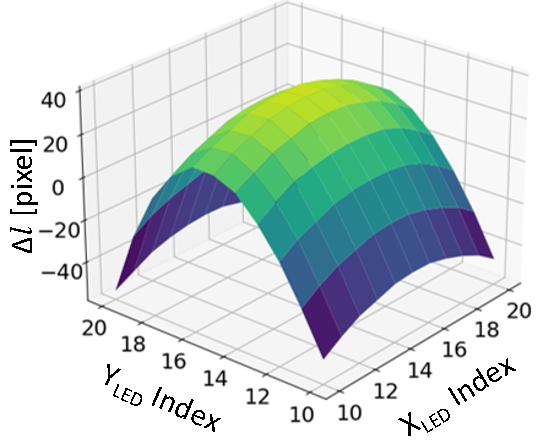}
      \caption{Minor Axes Lengths}
  \end{subfigure}
  \hfill
  \begin{subfigure}{0.22\textwidth}
      \includegraphics[width=\textwidth]{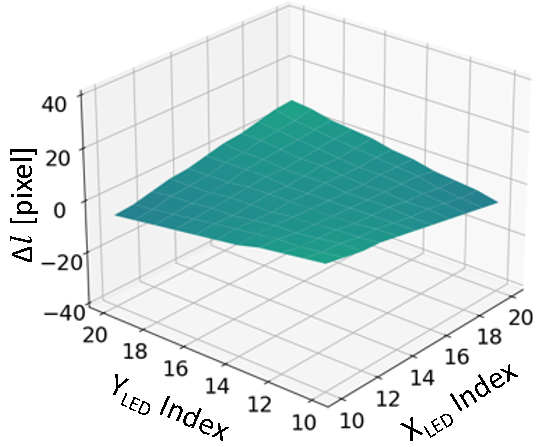}
      \caption{Major Axes Lengths}
  \end{subfigure}
  \caption{
  Visualization of the vignetting model parameters. Each data point in
  the grid corresponds to one LED being active, the plots show the
  respective ellipse parameters. In (c) and (d), $\Delta l$ represents the axes lengths subtracted from their corresponding linearlizations. }
  \label{fig:all_params}
\end{figure}

Experimentally, we instead observe varying elliptical shapes of the
vignetting circle, Fig.~\ref{fig:grid}, with non-linear translations
as evidenced by their centers plotted in
Fig.~\ref{fig:all_params}~(a) w.r.t. different LED positions. In the
following, we aim to explain this discrepancy.

We observe an elliptical shape of the vignetting circle for the center LED.
This shape can be explained by a
sensor tilt as the circular cone generated by the virtual point source
is intersected by a tilted plane, resulting in the observed
ellipse. Its minor axis corresponds to the tilt axis, whereas its
eccentricity is related to the tilt angle. Notably, a tilt in the LED
plane will not cause an elliptical shape as long as the exit pupil is
circular and parallel to the sensor plane.

To first order, an off-axis change of the LED position leads to a
linear change in the length of the ellipse axes, while their orientation
remains constant. We validated that our system parameters
$\sysvectgeom^*$, are in this linear range.  Due to the
observations in the previous paragraph, we expect this component to be
present in our system. In order to validate this hypothesis, we
individually fit ellipses to all accessible vignetting circles,
Fig.~\ref{fig:grid} (bottom, white outlines), followed by a fit of a
polynomial model in each of the ellipse parameters (center, axes,
orientation) across LED field positions. This model is refined on
partially visible vignetting circles, Fig.~\ref{fig:grid} (bottom,
blue outlines). Since we validated a sensor tilt in the first step, we
subtract the linear component of the ellipse's major and minor axes
model, in Fig.~\ref{fig:all_params}~(c) and (d) to underscore the
presence of higher order distortion terms in the optical system.

The shown residual components and the orientation field, Fig.~\ref{fig:all_params}~(b), of the
ellipses form a preliminary probe into the aberration behavior of the
optical system: Assuming that the ellipses result from a linear
stretch of the image along their major and minor axes, we can relate
this assumed geometric image distortion to a refocus term in the
wavefront. The varying ellipse components, excluding tilt-induced
changes, therefore indicate an LED field-varying astigmatism being
present in the system. In practice, this results in a combination of
blur and geometric distortion that should be compensated for optimal
performance.
We therefore recommend to use a full 4D aberration function to improve the reconstruction quality, 
however, it is beyond the scope of this work to demonstrate the expected effect.

The described LED-field
varying predictive model has the additional benefit of being usable to
predict bright- and dark-field regions in an image. For the full
benefit, the vignetting caused by the objective lens (second, unfitted
circular arc in Fig.~\ref{fig:grid} bottom) should also be incorporated.

\paragraph{Conclusions.}
In conclusion, we have investigated vignetting effects in FPM and
identified them as a rich source of information about the underlying
optical system.  A primary benefit of examining vignetting is its
critical role in optimizing the alignment of the optical system.
Moreover, these effects not only allow us to determine the geometric
construction parameters of the system, but also enable us to
characterize the partial coherence properties necessary to predict
interference effects faithfully in simulations.  Furthermore, our
analysis reveals that aberrations are evident in our setup,
necessitating an advanced 4D treatment. We expect that this approach
will yield substantial improvements in the reconstruction results.

\begin{backmatter}
\bmsection{Funding} German Research Foundation (DFG) under grants FOR 5336 (IH 114/2-1 and MO 2962/11-1).

\smallskip
\noindent
\bmsection{Disclosures} The authors declare no conflicts of interest.

\bmsection{Data availability} Data underlying the results presented in this paper are not publicly available at this time but may be obtained from the authors upon reasonable request.

\end{backmatter}

\bibliography{references}

\bibliographyfullrefs{references}

\end{document}